\documentclass[5p,twocolumn,times]{elsarticle}

\usepackage{lipsum}
\usepackage{lineno,hyperref}
\modulolinenumbers[5]
\usepackage[pdftex]{color}
\usepackage[font=footnotesize,labelfont=bf]{caption}
\usepackage[font=footnotesize,labelfont=bf]{subcaption}
\journal{Journal of \LaTeX\ Templates}

\usepackage{amssymb}

\biboptions{compress}

\usepackage[figuresright]{rotating}

\begin{document}

\begin{frontmatter}

\title{Spatial dynamics of synergistic coinfection in rock-paper-scissors models}

\address[1]{Institute for Biodiversity and Ecosystem
Dynamics, University of Amsterdam, Science Park 904, 1098 XH
Amsterdam, The Netherlands}
\address[2]{School of Science and Technology, Federal University of Rio Grande do Norte\\
Caixa Postal 1524, 59072-970, Natal, RN, Brazil}
\address[3]{ Edmond and Lily Safra International Institute of Neuroscience, Santos Dumont Institute,
Av Santos Dumont 1560, 59280-000, Macaiba, RN, Brazil}
\address[4]{Department of Computer Engineering and Automation, Federal University of Rio Grande do Norte, Av. Senador Salgado Filho 300, Natal, 59078-970, Brazil}

\author[1,2]{J. Menezes}  
\author[3,4]{E. Rangel}

\begin{abstract}
We investigate the spatial dynamics of two disease epidemics reaching a three-species cyclic model.
Regardless of their species, all individuals are susceptible to being infected with two different pathogens, which spread through person-to-person contact. The occurrence of coinfection leads to a synergistic increase in the risk of hosts dying due to complications from either disease.
Our stochastic simulations show that departed areas inhabited by hosts of a single pathogen arise from random initial conditions. The single-disease spatial domains are bordered by interfaces of coinfected hosts whose dynamics are curvature-driven. Our findings show that the coarsening dynamics of the interface network are controlled by the fluctuations of coinfection waves invading the single-disease territories. As the coinfection mortality grows, the dynamics of the interface network attain the scaling regime. We discover that organisms' infection risk is maximised if the coinfection increases the death due to disease in $30\%$, and minimised as the network dynamics reach the scaling regime, with species populations being maximum. Our conclusions may help ecologists understand the dynamics of epidemics and their impact on the stability of ecosystems.
\end{abstract}

\end{frontmatter}

\section{Introduction}

\label{sec1}
Spatial species distribution has been shown to be responsible for ecosystem formation and stability \cite{ecology}. Because of this, much attention has been dedicated to understanding how the organisms' spatial interactions affect biodiversity.
For example, researchers have discovered that 
species coexistence may become possible if organisms' competition happens locally, leading to the formation of departed spatial domains \cite{Coli,Allelopathy}. This has been observed, for example, in experiments with three strains of \textit{Escherichia coli}, whose cyclic dominance - described by the rock-paper-scissors game rules -
is not sufficient to ensure the species' persistence \cite{bacteria}. The same phenomenon has also been observed in other biological systems, like Californian coral reef invertebrates and lizards in the inner Coast Range of California, thus revealing the vital role space plays in promoting biodiversity \cite{lizards,coral,mobiliahigh}.

Scientists have shown that behavioural strategies 
can minimise the individual risk of being contaminated by a viral disease transmitted person-to-person \cite{social1,disease4,disease3,disease2}. For example, social distancing rules have been implemented worldwide, with individual and collective gains \cite{socialdist,soc,doi:10.1126/science.abc8881}. 
Furthermore, mobility restrictions have been shown efficient in decreasing the number of infected organisms and, consequently, minimising the social impact of epidemics on communities \cite{mr0,mr1,mr2,10.1371/journal.pone.0254403}. However, to guarantee the maximum efficiency of these measures in protecting against disease contamination, they are subject to adjustments if mutation alters the predominant pathogen causing the epidemics \cite{CAPAROGLU2021111246}.
The spatial organisation plasticity resulting from controlling the organisms' dispersal may be fundamental to the improve the efficiency of mitigation strategies against the disease with changing transmissibility and mortality \cite{plasticity2,Gene,mutating1,mutate2,plasticity1}.

In recent years, there has been a growing interest in studying the impact of epidemic spreading on individual fitness and species extinction risk \cite{epidemicbook, tanimoto}. The ongoing COVID-19 pandemic has prompted researchers from various fields to explore how self-preservation strategies employed by different species can inform human social interactions \cite{epidemicprocess,doi:10.1073/pnas.2007658117}.
Multiple studies have demonstrated that organisms can reduce their infection risk during local epidemic outbreaks by adopting specific behavioural strategies \cite{social1, disease4, disease3, disease2}. These findings have inspired the implementation of mitigation tactics such as social distancing and mobility restrictions worldwide, aimed at protecting individuals and communities during the COVID-19 pandemic \cite{socialdist, soc, doi:10.1126/science.abc8881, mr0, mr1, mr2, 10.1371/journal.pone.0254403}. Such strategies have also proven effective in reducing the infection risk of organisms in spatial games featuring cyclic dominance \cite{combination, jcomp}.
Furthermore, social distancing and mobility restrictions have also effectively reduced organisms' infection risk in spatial games with a cyclic dominance \cite{combination,jcomp}. Maximised benefits are obtained when organisms are apt to adjust the strategies when the disease mortality and transmission are changed by an eventual pathogen mutation \cite{plasticity,eplsick}.

Here, we study a cyclic model of three species whose dominance follows the rock-paper-scissors game rules, where organisms
face two simultaneous disease epidemics.
All individuals are susceptible to being infected with one of the pathogens or being coinfected - the diseases propagate person-to-person. Considering a synergistic increase in the mortality rate of the coinfected hosts, we explore the impact on pathogens' coexistence and the effects on population dynamics \cite{synergestic,synergestic2,synergestic3, synergestic4,synergestic5,synergestic8,synergestic10,virulence}. Our main goal is to decipher the complexity of the diseases' spatial distribution and understand how synergistic coinfection mortality may lead to the extinction of pathogens.
We aim to answer the questions: i) how do the pathogens 
are distributed in space?; ii) what are the conditions for both viruses to coexist?; iii) how do the viruses' spatial dynamics affect organisms' infection risk?; iv) how do the viruses' spatial interactions impact the species populations?

The outline of this paper is as follows: in Sec.~\ref{sec2}, we introduce our stochastic model and describe the simulations. The spatial patterns are investigated in 
Sec.~\ref{sec3}. The theoretical prediction for the scaling exponent of the coarsening dynamics is demonstrated in Sec.~\ref{sec4}. In Sec.~\ref{sec5}, we address the role of coinfection-catalysed mortality in disease coexistence. Also, in 
Sec.~\ref{sec6}, we study the effects of
the diseases' spatial distribution on organisms' infection risk and species densities. Finally, our conclusions and discussions are highlighted in Sec. \ref{sec7}.



\section{The stochastic model}
\label{sec2}

We investigate the stochastic rock-paper-scissors model, where three species outcompete, obeying the rules: scissors cut paper, paper wraps rock, and rock crushes scissors. The species are denoted by $i$, with $i=1,2,3$, and the indentification $i=i\,+\,3\,\alpha$, where $\alpha$ is an integer. We simulate the occurrence of two simultaneous epidemics whose pathogens responsible for the diseases are transmissible person-to-person. 
Organisms of every species are equally susceptible to contamination with both pathogens independently, with coinfection being possible at any time. 
Once infected, an individual becomes sick, which may cause death; in the case of the organism's health recovery, there is no immunity against reinfection. 
In the case of coinfection, the hosts' debilitation is accelerated, with a synergistic mortality rate increase.

\subsection{Simulations}

We create stochastic simulations to investigate the effects of the spreading of two distinct disease epidemics in a cyclic spatial game. Our algorithm follows the May-Leonard implementation, which does not consider a conservation law for the total number of individuals \cite{leonard}. The realisations run in square lattices with periodic boundary conditions; the maximum number of organisms is $\mathcal{N}$, the total number of grid points - at most, one individual is present at each grid site.

We define the density of individuals of species $i$, $\rho_i(t)$, with $i=1,2,3$, as the fraction of the lattice occupied by individuals of the species $i$ at time $t$:
\begin{equation}
\rho_i(t)=\frac{I_i(t)}{\mathcal{N}},
\end{equation}
where $I_i(t)$ is the total number of organisms of species $i$ at time $t$. 

To begin the simulation, we prepare random initial conditions by allocating each organism at 
a randomly chosen grid point.
We assume that the initial density is the same
for every species: $\rho_i \approx \mathcal{N}/3$, with $i=1,2,3$; the initial number of individuals is the maximum integer number that fits on the lattice, $I_i (t=0)\,\approx \,\mathcal{N}/3$, with $i=1,2,...,3$ - the remaining grid sites are left empty in the initial state. 
Furthermore, for every species, the initial ratio of individuals contaminated by each pathogen is $3\%$.

The notation $h_i$ specifies healthy individuals of species $i$, with $i=1,2,3$. Sick organisms with pathogens $1$ or $2$ are represented by $s^{1}_i$ and$s^{2}_i$, respectively; $s^{1,2}_i$ indicates coinfected individuals. The notation $i$ stands for all individuals, irrespective of illness or health. Using this notation, we describe the stochastic interactions as:
\begin{itemize}
\item 
Selection: $ i\ j \to i\ \otimes\,$, with $ j = i+1$, where $\otimes$ represents an empty space. After a selection interaction, the grid point previously occupied by the individual of species $i+1$ becomes an empty space.
\item
Reproduction: $ i\ \otimes \to i\ i\,$. A new individual of any species occupies an available empty space.
\item 
Mobility: $ i\ \odot \to \odot\ i\,$, where $\odot$ indicates either an individual of any species or an empty site. When moving, an organism switches position with another individual of any species or with an empty space.
\item 
Infection by disease $1$: $$s^{1}_i\ h_j \to s^{1}_i\ s^{1}_j\,$$ $$s^{1,2}_i\ h_j \to s^{1,2}_i\ s^{1}_j\,$$
$$s^{1}_i\ s^{2}_j \to s^{1}_i\ s^{1,2}_j\,$$ $$s^{1,2}_i\ s^{2}_j \to s^{1,2}_i\ s^{1,2}_j\,$$ with $i,j=1,2,3$. An organism of species $i$ infected with pathogen $1$ contaminated a healthy individual or a host of pathogen $2$.
\item 
Infection by disease $2$: $$s^{2}_i\ h_j \to s^{2}_i\ s^{2}_j\,$$ $$s^{1,2}_i\ h_j \to s^{1,2}_i\ s^{2}_j\,$$
$$s^{2}_i\ s^{1}_j \to s^{2}_i\ s^{1,2}_j\,$$ $$s^{1,2}_i\ s^{1}_j \to s^{1,2}_i\ s^{1,2}_j\,$$ with $i,j=1,2,3$. An individual of species $i$ transmits the pathogen $2$ to a healthy individual or a host of pathogen $1$.
\item 
Cure of disease $1$: $$s^{1}_i \to h_i\,$$ $$s^{1,2}_i \to s^{2}_i\,$$ An individual cured from the illness caused by pathogen $1$ but becomes susceptible to reinfection with the same virus.
\item 
Cure of disease $2$: $$s^{2}_i \to h_i\,$$ $$s^{1,2}_i \to s^{1}_i\,$$ An organism cured from disease $2$ may be become sick again after being cured.
\item 
Death caused by disease $1$: $$s^{1}_i \to \otimes\,$$ $$s^{1,2}_i \to \otimes\,$$ When an ill individual dies because of the disease $1$, the grid site becomes empty.
\item
Death due to disease $2$: $$s^{2}_i \to \otimes\,$$ $$s^{1,2}_i \to \otimes\,$$ A sick individual infected with pathogen $2$ may die, leaving its grid point empty.
\end{itemize}

The interaction stochasticity is defined by a set of probabilities, which is computed according to the parameters: $S$ (reproduction), $R$ (reproduction),
$M$ (mobility), $\kappa$ (pathogen transmissibility), $\mu$ (disease mortality), and $C$ (cure). 
In case of
coinfected organism, pathogens transmission and mortality are rescaled by the factors $\epsilon$ and $\gamma$, respectively.

The actions are implemented using the Moore neighbourhood, where each organism may interact with one of its eight immediate neighbours. The algorithm proceeds as follows: i) randomly picking an active individual among all organisms in the lattice; ii) randomly choosing one interaction to be executed; iii) drawing one of the eight immediate neighbours to receive the action. 
One time step is counted if either an interaction is implemented. The time necessary for $\mathcal{N}$ to occur is defined as one generation, our time unit.

\section{Spatial patterns}
\label{sec3}
We first investigate the spatial patterns arising from random initial conditions in the rock-paper-scissors model with two contagious diseases, considering a synergistic increase in the mortality rate of coinfected rates.
For this purpose, we ran a single simulation in a lattice with $300^2$ starting from initial conditions for a timespan of $10000$ generations, with 
$S=R=1.0$, $M=3.0$, $\kappa=4.0$, $\mu=C=\epsilon=0.1$, and $\gamma=2.0$. This means coinfected individuals are twice as likely to die due to disease complications than sick organisms contaminated with a single disease.

\begin{figure*}[h]
	\centering
	  \begin{subfigure}{.19\textwidth}
        \centering
        \includegraphics[width=34mm]{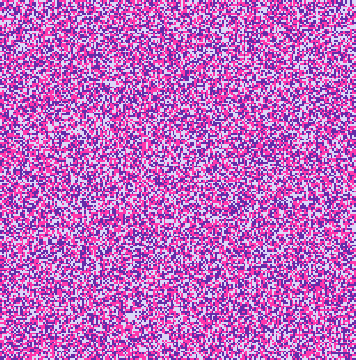}
        \caption{}\label{fig1a}
    \end{subfigure} 
    \begin{subfigure}{.19\textwidth}
        \centering
        \includegraphics[width=34mm]{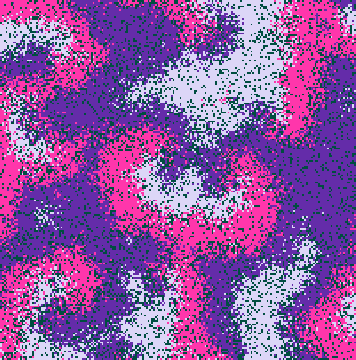}
        \caption{}\label{fig1b}
    \end{subfigure} %
   \begin{subfigure}{.19\textwidth}
        \centering
        \includegraphics[width=34mm]{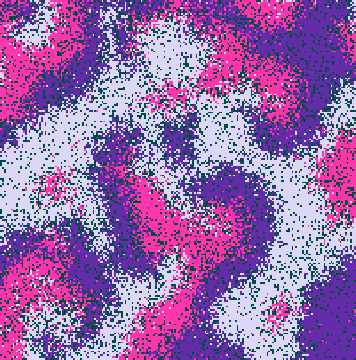}
        \caption{}\label{fig1c}
    \end{subfigure} 
            \begin{subfigure}{.19\textwidth}
        \centering
        \includegraphics[width=34mm]{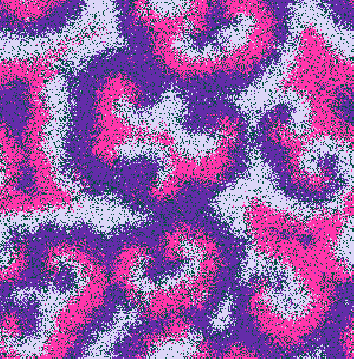}
        \caption{}\label{fig1d}
    \end{subfigure} 
           \begin{subfigure}{.19\textwidth}
        \centering
        \includegraphics[width=34mm]{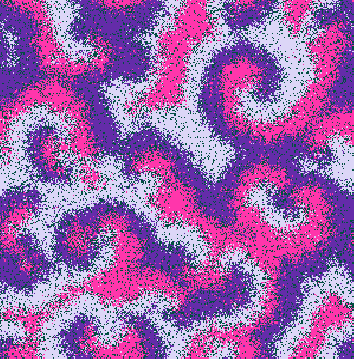}
        \caption{}\label{fig1e}
    \end{subfigure} \\
                \begin{subfigure}{.19\textwidth}
        \centering
        \includegraphics[width=34mm]{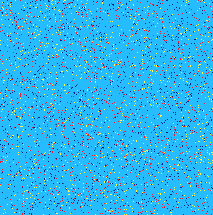}
        \caption{}\label{fig1f}
    \end{subfigure} %
   \begin{subfigure}{.19\textwidth}
        \centering
        \includegraphics[width=34mm]{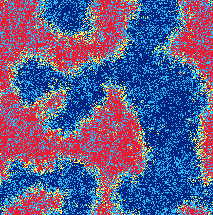}
        \caption{}\label{fig1g}
    \end{subfigure} 
            \begin{subfigure}{.19\textwidth}
        \centering
        \includegraphics[width=34mm]{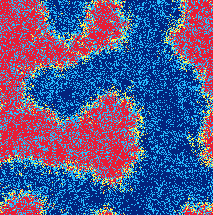}
        \caption{}\label{fig1h}
    \end{subfigure} 
           \begin{subfigure}{.19\textwidth}
        \centering
        \includegraphics[width=34mm]{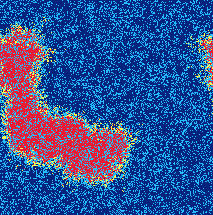}
        \caption{}\label{fig1i}
    \end{subfigure} 
   \begin{subfigure}{.19\textwidth}
        \centering
        \includegraphics[width=34mm]{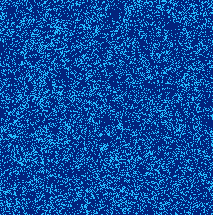}
        \caption{}\label{fig1j}
            \end{subfigure} 
            \begin{subfigure}{.19\textwidth}
                \centering
        \includegraphics[width=34mm]{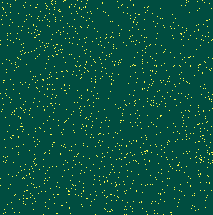}
        \caption{}\label{fig1k}
    \end{subfigure} 
            \begin{subfigure}{.19\textwidth}
        \centering
        \includegraphics[width=34mm]{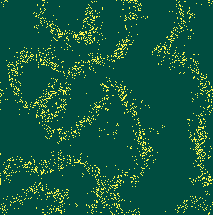}
        \caption{}\label{fig1l}
    \end{subfigure} 
           \begin{subfigure}{.19\textwidth}
        \centering
        \includegraphics[width=34mm]{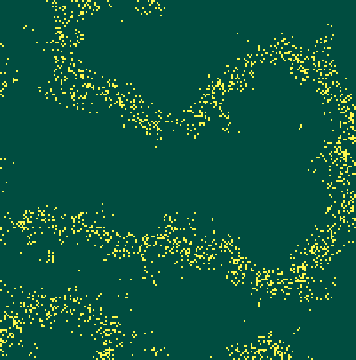}
        \caption{}\label{fig1m}
    \end{subfigure} 
   \begin{subfigure}{.19\textwidth}
        \centering
        \includegraphics[width=34mm]{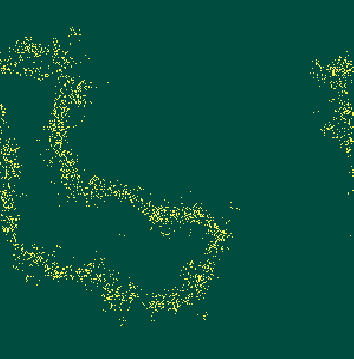}
        \caption{}\label{fig1n}
            \end{subfigure}
               \begin{subfigure}{.19\textwidth}
        \centering
        \includegraphics[width=34mm]{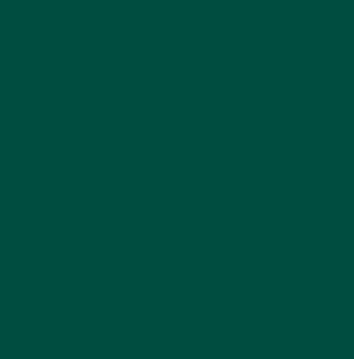}
        \caption{}\label{fig1o}
            \end{subfigure}
 \caption{Snapshots of the rock-paper-scissors model with two disease epidemics. 
Figures \ref{fig1a} to \ref{fig1e} show the organism's configuration at the initial conditions and after $600$, $1200$, $4000$, and $9600$ generations; pink, dark purple, and light purple dots show individuals of species $1$, $2$, and $3$, respectively; green dots depict empty spaces. Red and dark blue dots in Figs. \ref{fig1f} to \ref{fig1j} show the respective spatial distribution of individuals infected with virus $1$ and $2$; yellow dots show the coinfected organisms while light blue dots are healthy individuals and empty spaces. Figures \ref{fig1k} to \ref{fig1o} highlight the interfaces of coinfected organisms in yellow, the spatial positions depicted in dark green are empty, occupied by single-disease or healthy individuals.
}
  \label{fig1}
\end{figure*}

To observe what happens during the simulation, we prepare three visualisations as follows:
\begin{itemize}
\item 
Figures \ref{fig1a} to \ref{fig1e} show the organisms of each species, irrespective of whether they are healthy, infected with a single pathogen or coinfected. 
The colours pink, dark purple, and light purple
identify the species $1$, $2$, $3$, respectively; green dots show the empty spaces. 
\item
Figures \ref{fig1f} to \ref{fig1j} depict the spatial distribution of organisms infected with pathogen $1$ (red), pathogen $2$ (dark blue), and coinfected (yellow); light blue dots show healthy individuals and empty spaces.
\item
Figures \ref{fig1k} to \ref{fig1o} highlight the grid sites with coinfected hosts, which are depicted in yellow; green areas show the absence of coinfection (single infected, healthy individuals and empty spaces).
\end{itemize}

Videos {https://youtu.be/k0LmZQmQtYg}, {https://youtu.be/hBsgxrjBU1U}, and {https://youtu.be/oujTTV9f3kU} show the dynamics of the organisms' spatial organisation, the disease distribution, and coinfected hosts, respectively, during the whole simulation. The snapshots in Fig.~\ref{fig1} were captured of the simulation at $t=0$ (Figs. \ref{fig1a}, \ref{fig1f}, and \ref{fig1k}), $t=600$ (Figs. \ref{fig1b}, \ref{fig1g}, and \ref{fig1l}), $t=1200$ (Figs. \ref{fig1c}, \ref{fig1h}, and \ref{fig1m}), $t=4000$ (Figs. \ref{fig1d}, \ref{fig1i}, and \ref{fig1n}), and $t=9600$ (Figs. \ref{fig1e}, \ref{fig1j}, and \ref{fig1o}).

As we observe in Figs.~\ref{fig1a} to \ref{fig1e}, individuals of the same species occupy departed spatial domains after an initial stage of the pattern formation process. Because of the cyclic selection rules of the rock-paper-scissors game, spirals emerge from the random initial conditions. The outcomes show
the spreading of both pathogens that initially separately infest only $3\%$ of individuals, thus leading to the coinfection of many hosts.

Since single-pathogen hosts survive longer than coinfected ones, there is a spontaneous arising of two kinds of spatial domains mostly occupied by sick individuals with disease $1$ (red) or illness $2$ (dark blue). Because of the spatial topology, 
coinfected individuals (yellow) concentrate in the interfaces on the borders between red and dark blue regions, as shown in Fig.~\ref{fig1g} and highlighted in 
Fig.~\ref{fig1l}. 
As time passes, the total length of the interfaces of coinfected organisms reduces. Figures ~\ref{fig1h} to ~\ref{fig1j} show that the straightening minimises the size of the red area, resulting in its collapse. Therefore, the topological features of the two-dimensional space induce the extinction of one of the pathogens. The dynamics of the interface of coinfected individuals leading to the end of one disease are shown by the reduction and further disappearance of the yellow dots in Figs. ~\ref{fig1m} to ~\ref{fig1o}.

\subsection{Dynamics of species densities}

\begin{figure}[t]
\centering
\includegraphics[width=90mm]{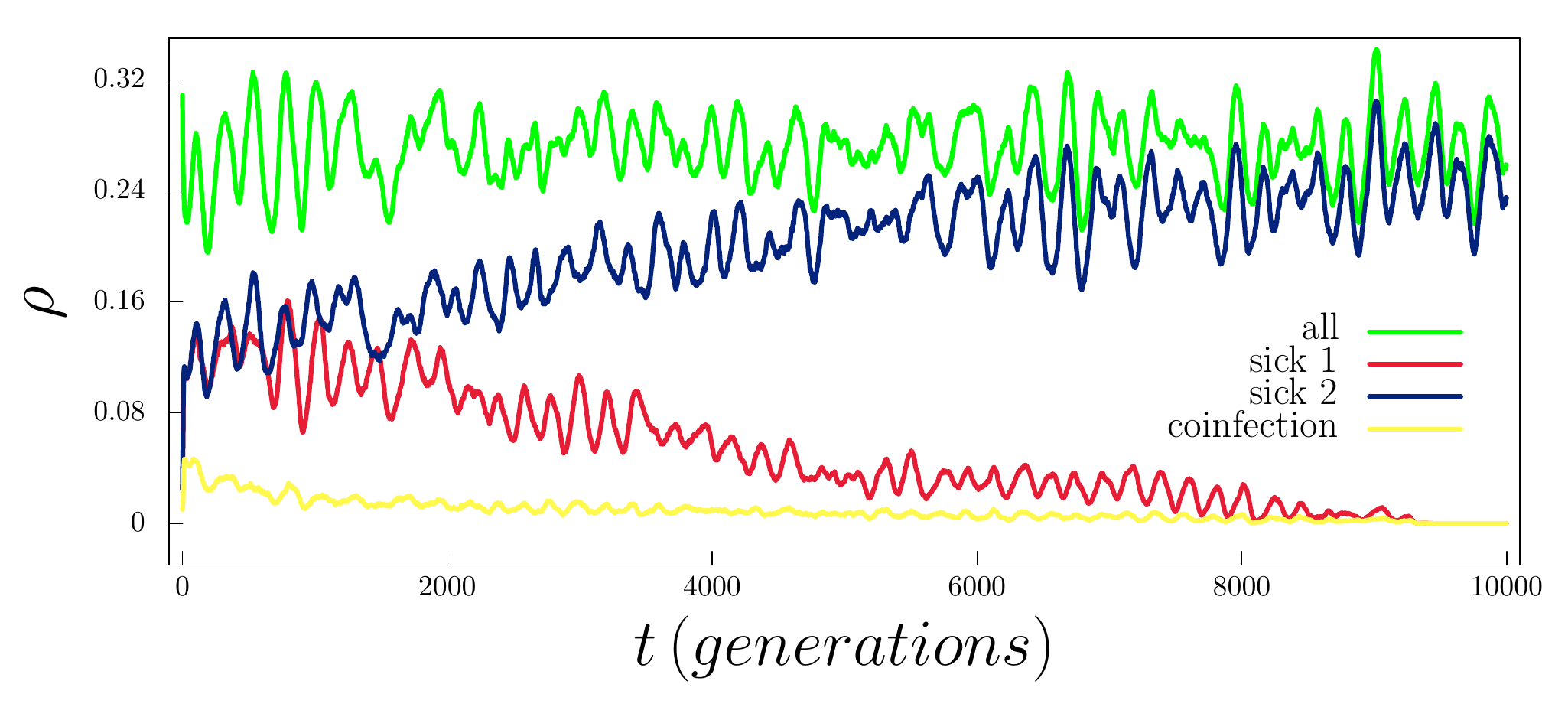}
\caption{Dynamics of species densities in the simulation shown in Fig.~\ref{fig1}. The green line shows how the species density of individuals of species $1$ varies with time, irrespective of being healthy or sick. The red, dark blue, and yellow lines depict the temporal dependence of the density of organisms
of species $1$ infected with virus $1$, virus $2$ and coinfected, respectively.}	
\label{fig2}
\end{figure}

We also calculated the temporal dependence of the densities of organisms of each species during the entire simulation. The outcomes are shown in Fig.~\ref{fig2}, where the green line shows the dynamics of the density of individuals of species $1$. The red and dark blue lines depict the density of individuals of species $1$ contaminated with disease $1$ and $2$, respectively; the yellow line shows the density of coinfected individuals of species $1$. Our choice for presenting the outcomes of species $1$ is arbitrary since the symmetry of the rock-paper-scissors game with the assumption of the same parameters for all species results in the same average densities for every species.

Figure \ref{fig2} shows two independent dynamics related to the cyclic spatial games and the epidemics. According to the green line, the temporal dependence of the species densities follows the symmetry inherent to the rock-paper-scissors game: alternating dominance with constant average density. 
Concerning the densities of hosts of each pathogen, 
the results show that the synergistic effects of coinfection provoke the increase in the average density of sick individuals infected with the virus $1$ (red) and, consequently, the reduction in the proportion of individuals of each species contaminated with the virus $2$ (dark blue). In addition, the yellow line depicts the decline in the density of coinfected individuals as the total length of the interface decreases.
\section{Scaling Regime} \label{sec4}

In the previous section, we found that single-pathogen domains surrounded by interfaces mainly composed of coinfected organisms arise from random initial conditions.
The interface network is similar to those which appear in systems of two species which directly compete for space or whose competition is apparent - induced by the presence of a common predator \cite{Avelino-PRE-86-031119,Pereira,apparent}.

In our model, simultaneously contaminating an organism, there is a synergistic increase in disease severity. Therefore, the acceleration in coinfected organisms' health depletion is the agent that breaks the spatial symmetry of the disease distribution, leading to each pathogen dominating departed areas. 
This means that the extinction of one of the diseases results from a coarsening process which leads some territories occupied by a single virus to grow, provoking the collapse of other single-disease domains \cite{Avelino-PRE-86-031119,apparent}.

Based on the topological aspects of two-dimensional space, we investigate the dynamics of pathogens' spatial distribution considering an interface with curvature radius $r_{\kappa}$ and thickness $\varepsilon\,\ll\,r_{\kappa}$, separating two distinct single-prey domains. Suppose organisms infected with virus $1$ are outside the interface while disease provoked by pathogen $2$ occupies the inner region.
The interface comprises coinfected organisms  
transmitting the diseases in both directions, reaching immediate organisms outside and inside the interface. 
Thus, the transmission of pathogen $2$ can affect a number of hosts of pathogen $1$ that is proportional to the outer interface length: $r_{\kappa} + \varepsilon/2$. Also, the number of individuals vulnerable to being infected with $1$ is proportional
to the inner interface length, which is proportional to $r_{\kappa} - \varepsilon/2$.

As the transmission of pathogens $1$ and $2$ occurs at the same rate, the difference between the average number of newly coinfected organisms per unit of times from outside and inside the interface is proportional to $\varepsilon$. This means that the time necessary for all organisms to become coinfected
inside the interface is shorter than for all individuals outside the interface. The consequence is that,
as time passes, one has a reduction of the interface curvature radius and the consequent collapse of closed interfaces.

\begin{figure*}[h]
	\centering
    \begin{subfigure}{.19\textwidth}
        \centering
        \includegraphics[width=34mm]{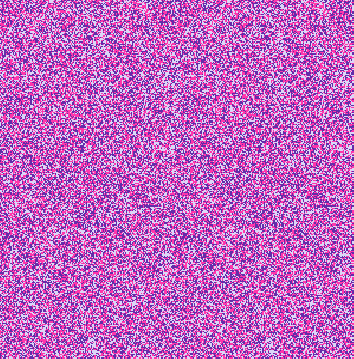}
        \caption{}\label{fig3a}
    \end{subfigure} %
   \begin{subfigure}{.19\textwidth}
        \centering
        \includegraphics[width=34mm]{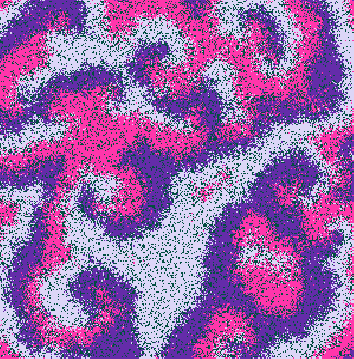}
        \caption{}\label{fig3b}
    \end{subfigure} 
            \begin{subfigure}{.19\textwidth}
        \centering
        \includegraphics[width=34mm]{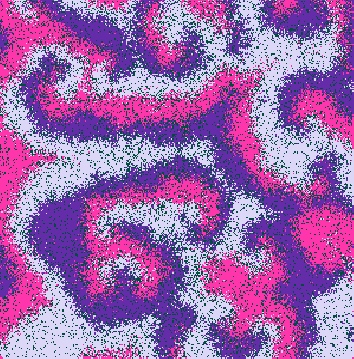}
        \caption{}\label{fig3c}
    \end{subfigure} 
           \begin{subfigure}{.19\textwidth}
        \centering
        \includegraphics[width=34mm]{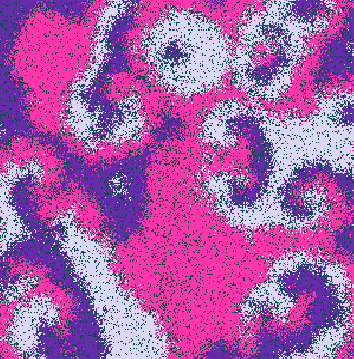}
        \caption{}\label{fig3d}
    \end{subfigure} 
   \begin{subfigure}{.19\textwidth}
        \centering
        \includegraphics[width=34mm]{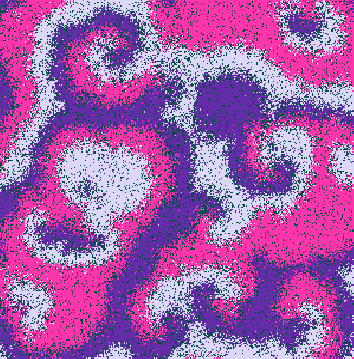}
        \caption{}\label{fig3e}
            \end{subfigure}\\
                \begin{subfigure}{.19\textwidth}
        \centering
        \includegraphics[width=34mm]{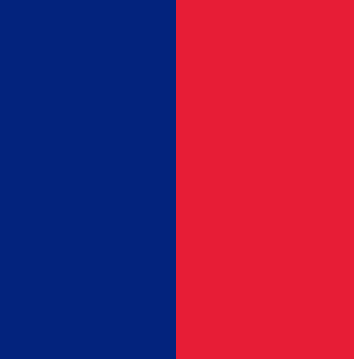}
        \caption{}\label{fig3f}
    \end{subfigure} %
   \begin{subfigure}{.19\textwidth}
        \centering
        \includegraphics[width=34mm]{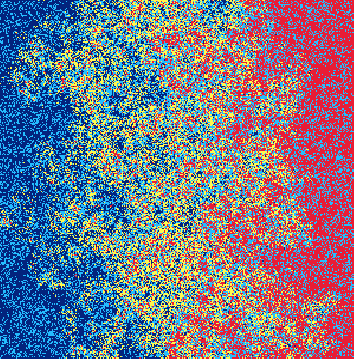}
        \caption{}\label{fig3g}
    \end{subfigure} 
            \begin{subfigure}{.19\textwidth}
        \centering
        \includegraphics[width=34mm]{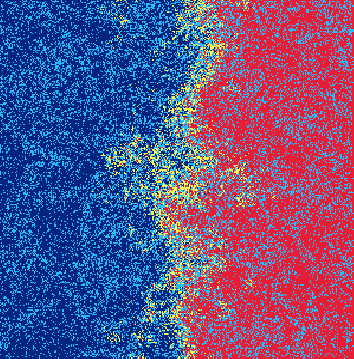}
        \caption{}\label{fig3h}
    \end{subfigure} 
           \begin{subfigure}{.19\textwidth}
        \centering
        \includegraphics[width=34mm]{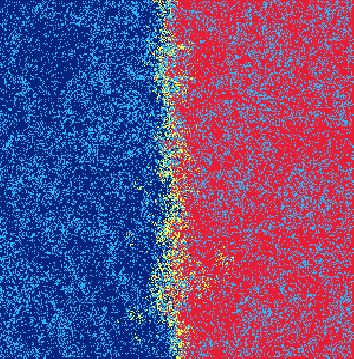}
        \caption{}\label{fig3i}
    \end{subfigure} 
   \begin{subfigure}{.19\textwidth}
        \centering
        \includegraphics[width=34mm]{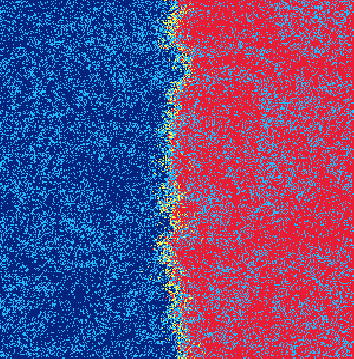}
        \caption{}\label{fig3j}
            \end{subfigure}
        \begin{subfigure}{.19\textwidth}
        \centering
        \includegraphics[width=34mm]{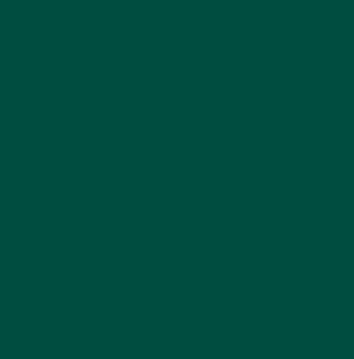}
        \caption{}\label{fig3k}
    \end{subfigure} %
   \begin{subfigure}{.19\textwidth}
        \centering
        \includegraphics[width=34mm]{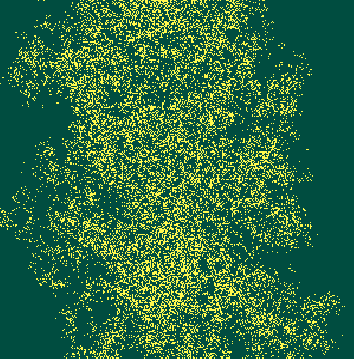}
        \caption{}\label{fig3l}
    \end{subfigure} 
            \begin{subfigure}{.19\textwidth}
        \centering
        \includegraphics[width=34mm]{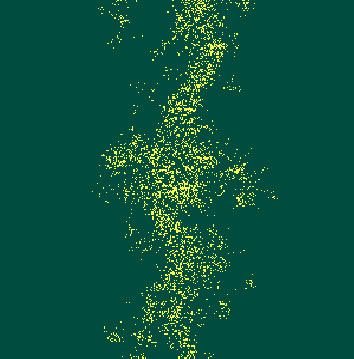}
        \caption{}\label{fig3m}
    \end{subfigure} 
           \begin{subfigure}{.19\textwidth}
        \centering
        \includegraphics[width=34mm]{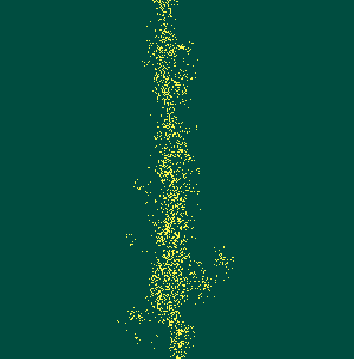}
        \caption{}\label{fig3n}
    \end{subfigure} 
   \begin{subfigure}{.19\textwidth}
        \centering
        \includegraphics[width=34mm]{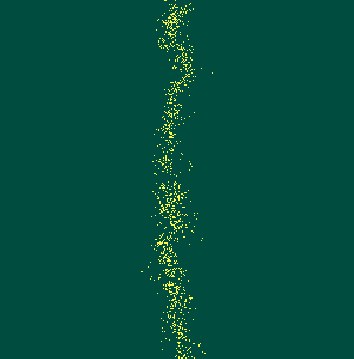}
        \caption{}\label{fig3o}
            \end{subfigure}
 \caption{Snapshots of simulations of a single interface of coinfected individuals for various coinfection mortality catalyst factors.
Figures \ref{fig3a}, \ref{fig3f}, and \ref{fig3k} shows
the initial prepared organisms' organisation, spatial disease distribution and absence of coinfected; the colours follow the same scheme in Fig. \ref{fig1}.
The spatial configuration after $800$ generations 
appears in Figs. \ref{fig3b}, \ref{fig3g} and  \ref{fig3l} for $\gamma=1.25$; the results for $\gamma=1.5$ are shown in Figs. \ref{fig3c}, \ref{fig3h} and  \ref{fig3m}. For higher $\kappa$, the interface fluctuations reduces, as shown in
Figs. \ref{fig3d}, \ref{fig3i} and  \ref{fig3n} for $\gamma=1.75$, and Figs. \ref{fig3e}, \ref{fig3j} and  \ref{fig3o} for $\gamma=2.0$.}
  \label{fig3}
\end{figure*}

We then conclude that the time necessary for the radius of a circular single-disease area reduces by $\Delta r_{\kappa} \ll r_{\kappa}$ is proportional 
to the interface length, i.e., proportional to $r_{\kappa}$. Hence, the dynamics of the interface network are curvature-driven: the speed of the collapse of single-pathogen spatial domains is proportional to its curvature, which is inherent to non-relativistic interfaces in condensed matter \cite{Joana1,Joana2}.

\subsection{Scaling power law}

The characteristic length of the coinfection interface network at time $t$ using is defined as
\begin{equation} 
L(t)\,=\,\sqrt{\frac{A}{N_D(t)}}\,=\,\sqrt{\frac{\mathcal{N}}{N_D(t)}} \label{eeq1}
\end{equation} 
where $A$ is the total grid area (the total number of grid points $\mathcal{N}$) and $N_{D} (t)$ is the number of single-pathogen domains in the time $t$. 
As the dynamics of the interface network are curvature-driven, the number of single-pathogen spatial domains decreases as time passes \cite{Avelino-PRE-86-031119,apparent}: 
\begin{equation}
N_D(t)\,=\, \frac{G}{t} \label{eqq1}
\end{equation}
where $G$ is a positive constant.
Hence,
the characteristic length
grows following the scaling law
\begin{equation}
L\,\propto t^{1/2}, \label{eq1}
\end{equation}
which describes the dynamics of the coinfection interface networks (yellow dots in Figs.~\ref{fig1l} to \ref{fig1m}).

Furthermore, using the notation $I^{\star}$ to represent the total number of coinfected individuals,
one has 
$I^{\star}\,=\,\varepsilon\,L_T$
where $L_T$ is total interface length and $\varepsilon$ the interface thickness. As the average interface thickness is constant in time and space (since the parameters used in the stochastic simulations do not vary in space and time \cite{Roman}), one has
\begin{equation}
I^{\star}\,\propto\, L_T. \label{eqt}
\end{equation}

\begin{figure}[t]
\centering
        \includegraphics[width=75mm]{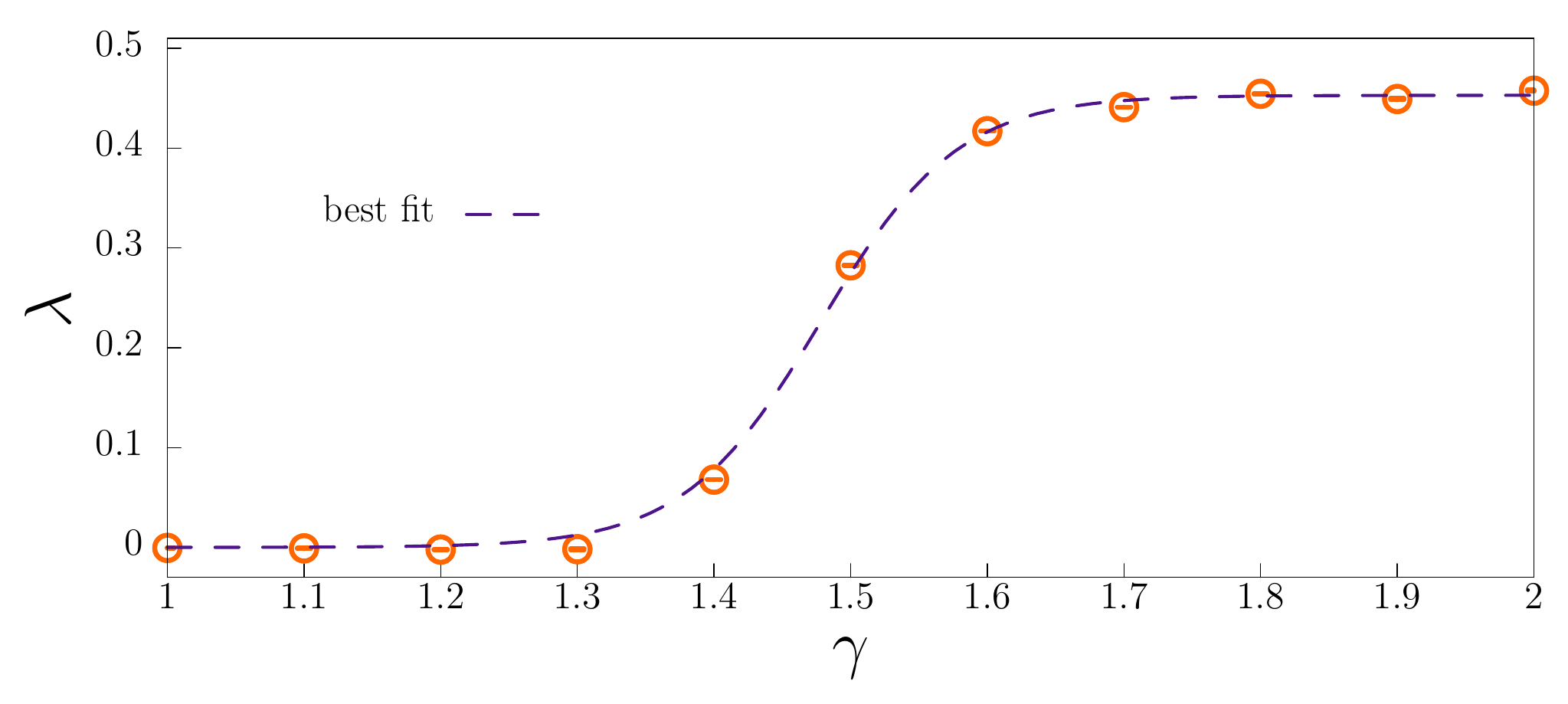}
\caption{Scaling exponent in terms of the coinfection catalysed mortality. The orange dots show
the $\lambda$ computed using the function $I^{\star} \propto t^{\lambda}$, where 
$I^{\star}$ is the number of coinfected individuals.
The outcomes were averaged from $100$ simulations running in lattices with $300^2$ grid sites for a timespan of $10000$ generations; the error bars indicate the standard deviation. 
The dashed purple line shows the best fit of the numerical results.}
	\label{fig4}
\end{figure}

We also notice that the number of single-prey domains is given by
\begin{equation}
N_D\,=\,\frac{L_T}{P_D}, \label{eqr}
\end{equation}
where $P_D$ is the average domain perimeter. 
Given that the average domain radius defines the network characteristic length $L$, one has 
\begin{equation}
P_D\,\propto\, L, \label{eqe}
\end{equation}
which allows Eq.~\ref{eqr} to be written as 
\begin{equation}
N_D\,\propto\,\frac{L_T}{L}. \label{eqree}
\end{equation}  

Combining Eqs.~\ref{eeq1} and
~\ref{eqree}, one has
\begin{equation}
I^{\star}\,\propto\,L^{-1}. \label{eqi}
\end{equation}
Finally, substituting Eq.~\ref{eq1} in Eq.~\ref{eqi}, we find the scaling power law that characterises the dynamics of the interface networks in our two-pathogen model competition model as
\begin{equation}
I^{\star}\,\propto\,t^{-1/2}. \label{eqo}
\end{equation}

We verified the analytical prediction of
the scaling law given by Eq.~\ref{eqo}, using the outcomes of a set of $100$ simulations starting from different initial conditions. The realisations ran
in lattices with $300^2$ sites, with a timespan of $10000$ generations - the parameters are the same as in the simulation in Fig.~\ref{fig1}. 
Given that each simulation started from different random initial conditions, we computed the average scaling exponent $\lambda$ fitting the dynamics of the total number of coinfected individuals using the function
 $I^{\star} \propto t^{-\lambda}$. We have not considered the first $150$ generations to calculate the scaling exponent to avoid the transient interface network formation fluctuations observed in Fig.\ref{fig2}.
We found that $\lambda = 0.4576 \pm 0.00036$, close to the theoretical prediction in Eq.~\ref{eqo}. This confirms that the dynamics of the coinfected interface network approach the expected scaling regime; the deviation is due to grid size finiteness \cite{Avelino-PRE-86-031119}.
 
\section{The impact of synergistic mortality on the coarsening dynamics} \label{sec5}
We now study the influence of the synergistic mortality factor, $\gamma$, on the pattern formation process and the coarsening dynamics of the single-pathogen domains. For this purpose, we prepare the initial conditions shown in Figs.~\ref{fig3a}, ~\ref{fig3f} and ~\ref{fig3k}: i) organisms of every species are randomly allocated on the lattice (the number of individuals of each is the same); ii) on the left side,
all individuals host the pathogen $1$ (red); on the right side, all organisms are infected with pathogen $2$ (red); iii) there is no coinfected individual in the initial conditions.

We performed four simulations assuming different synergistic mortality factors and captured the spatial configuration after $1000$ generations: i) $\gamma=1.25$: Figs.~\ref{fig3b}, ~\ref{fig3g}, and ~\ref{fig3l}; ii) $\gamma=1.5$: Figs.~\ref{fig3c}, ~\ref{fig3h}, and ~\ref{fig3m}; iii) $\gamma=1.75$: Figs.~\ref{fig3d}, ~\ref{fig3i}, and ~\ref{fig3n}; iv) $\gamma=2.0$: Figs.~\ref{fig3e}, ~\ref{fig3j}, and ~\ref{fig3o}. We used lattices with $300^2$ grid sites; the other parameters are the same as in the previous simulations. We relaxed the periodic boundary to observe the dynamics of a central vertical interface, assuming a lattice with fixed edges.                 

As in Figure \ref{fig1}, the spatial patterns show spiral waves whose dynamics are controlled by rock-paper-scissors rules, not being relevantly affected by the mortality acceleration of coinfected individuals, as one observes in Figs.~\ref{fig1b} to \ref{fig1e}. On the other hand, the level of mortality of coinfected hosts controls the stability of the central interface. In summary,
\begin{itemize}
\item 
Organisms on the border are easily infected by the virus spreading from the opposite spatial domain. Thus, an interface of yellow dots appears after the simulation begins.
\item
For low $\gamma=1.25$, coinfected individuals continue passing the pathogen $1$ to organisms with the dark blue side; the same happens on the opposite side where pathogen $2$ continues advancing. After $800$ generations, almost all individuals in the grid are coinfected, as shown in the snapshots in Figs.~\ref{fig3g} and \ref{fig3l}. 
\item
As $\gamma$ grows, the chances of organisms that become coinfected on the border moving toward the interior of the red and dark blue regions reduce. This happens because a fraction of them die before transmitting the pathogens. Figures \ref{fig3h} and \ref{fig3m} show that for $\gamma=1.5$, coinfection waves stochastically propagate toward single-disease regions but disappear when organisms die. For $\gamma=1.75$, the probability of coinfection reaching individuals distant from the interface reduces, as shown in Figs. \ref{fig3i} and \ref{fig3n}.
\item
Finally, the outcomes show that for $\gamma=2.0$ - where mortality is twice likely for coinfected hosts - the chances of the virus spreading in the opposite spatial domains significantly reduce. 
In this case, coinfection waves quickly stop propagating into the single-disease domains due to the high probability of death of coinfected individuals.
\end{itemize}

We then quantify how the propagation of coinfection waves into the single-disease domains interferes with the coarsening dynamics predicted in the previous section. To achieve this goal, we ran sets of $100$ simulations for various values of $\gamma$ starting from random initial conditions. 
The simulations were performed in lattices with $300^2$ sites, running until $6000$ generations, for $1\,\leq\,\gamma\,\leq 2$, in intervals of $\Delta \gamma =0.1$.

To measure the scaling exponent, we assume a generalised version of Eq.~\ref{eqo} as
\begin{equation}
I^{\star} \propto t^{-\lambda(\gamma)}, \label{eqcross}
\end{equation}
where $\lambda(\sigma)$ is a function of the synergistic mortality factor.

Figure~\ref{fig4} shows the average scaling exponent $\lambda$ value, where the error bars show the standard deviations.
The results confirm that for low $\gamma$, the coarsening dynamics depart from a curvature-driven scenario, with the scaling exponent indicating that the pathogens coexist. For $\gamma <1.2$, our findings show that the number of coinfected individuals is approximately constant in time.
In contrast, for $\gamma >1.8$, the coarsening dynamics approaches the scaling regime predicted in Eq. \ref{eqo} since the high mortality rate of coinfected individuals reduces the impact of the fluctuations.

Therefore, as the coinfection becomes deadly ($\gamma$ grows), there is a crossover from the coexistence regime ($\lambda \approx 0.0$) and the interface network coarsening ($\lambda \approx 0.5$). We found the best fit to describe the dependence of the scaling exponent $\lambda$ on the synergistic mortality factor:
\begin{equation}
\lambda (\gamma) = A\,\left[ 1\,-\tanh (B-10\, \gamma) \right], \nonumber
\end{equation}
where $A=0.226\, (\pm 0.001)$ and $B=14.78\, (\pm 0.018)$, 
as depicted by the dashed purple line in Fig.~\ref{fig4}.

\begin{figure}
\centering
        \includegraphics[width=75mm]{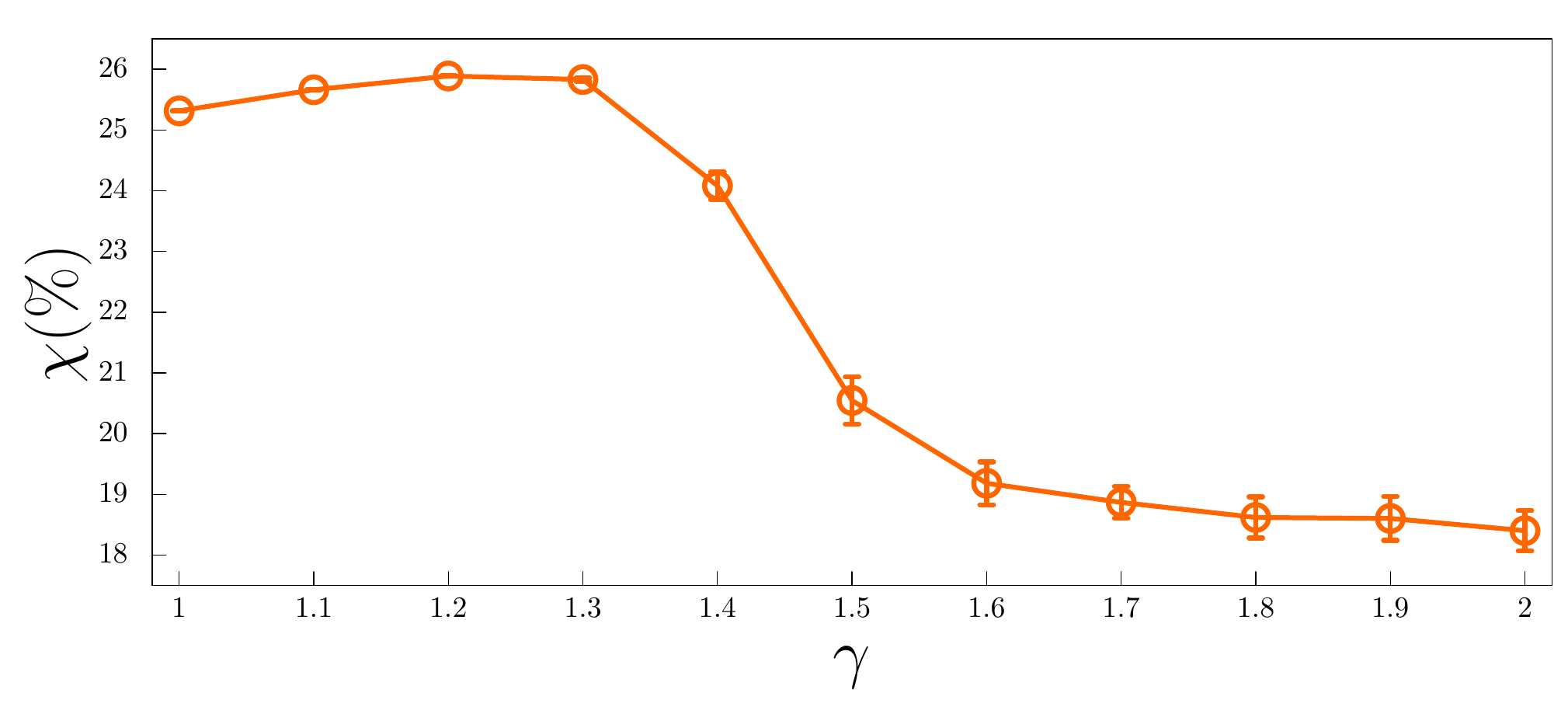}
\caption{Organisms' infection risk as a function of the coinfection mortality catalysed factor. The results were averaged from sets of $100$ simulations running in lattices with $500^2$ grid sites, running until $t=5000$ generations. The error bars show the standard deviation.}
	\label{fig5}
\end{figure}

We then conclude that the higher the coinfection mortality is, the less the dynamics of the spatial patterns are affected by the noise introduced by waves spreading through the single-pathogen species observed in Fig. ~\ref{fig3}. This is similar to the thermal fluctuations that roughen the interfaces, thus opposing the curvature-driven growth and decelerating the two-dimensional coarsening dynamics \cite{Topics, Arezon1,Interplay,Arezon2,Oliveira1,Tartaglia_2018}.

\section{Influence of synergistic mortality at individual and population levels}
\label{sec6}
Finally, we investigate the influence of synergistic mortality factors on organisms' infection risk and population dynamics. For this reason, we ran sets of $100$ simulations starting from different initial conditions in lattices with $500^2$ grid lattices running until $5000$ generations. We conduct simulations for $0\leq \gamma \leq 1$, in intervals of $\gamma=0.1$; the other parameters are the same as in the previous sections.

We define the infection risk, $\chi(t)$, as the probability of a healthy individual being infected by one of the pathogens per unit of time, respectively. For this purpose, the algorithm proceeds as follows:
i) counting the total number of healthy individuals of species $i$ when each generation begins; ii) computing the number of healthy individuals of species $i$ that are contaminated (virus $1$ or virus $2$) during the generation; 
iii) calculating the infection risk $\chi_i$, with $i=1,2,3$, defined as the ratio between the number of infected individuals and the initial number of healthy individuals of species $i$.

In addition, we compute
the average value of the density of organisms of species $i$, $\rho_1$.
Due to the symmetry of the rock-paper-scissors game, the results are the same for every species; thus, we use the data from species $1$ to compute $\chi$ and $\rho$. The statistical analyses avoided the high fluctuations inherent to the pattern formation process: we calculate the mean infection risk and species densities using the data from the second simulation half.

\begin{figure}
\centering
        \includegraphics[width=75mm]{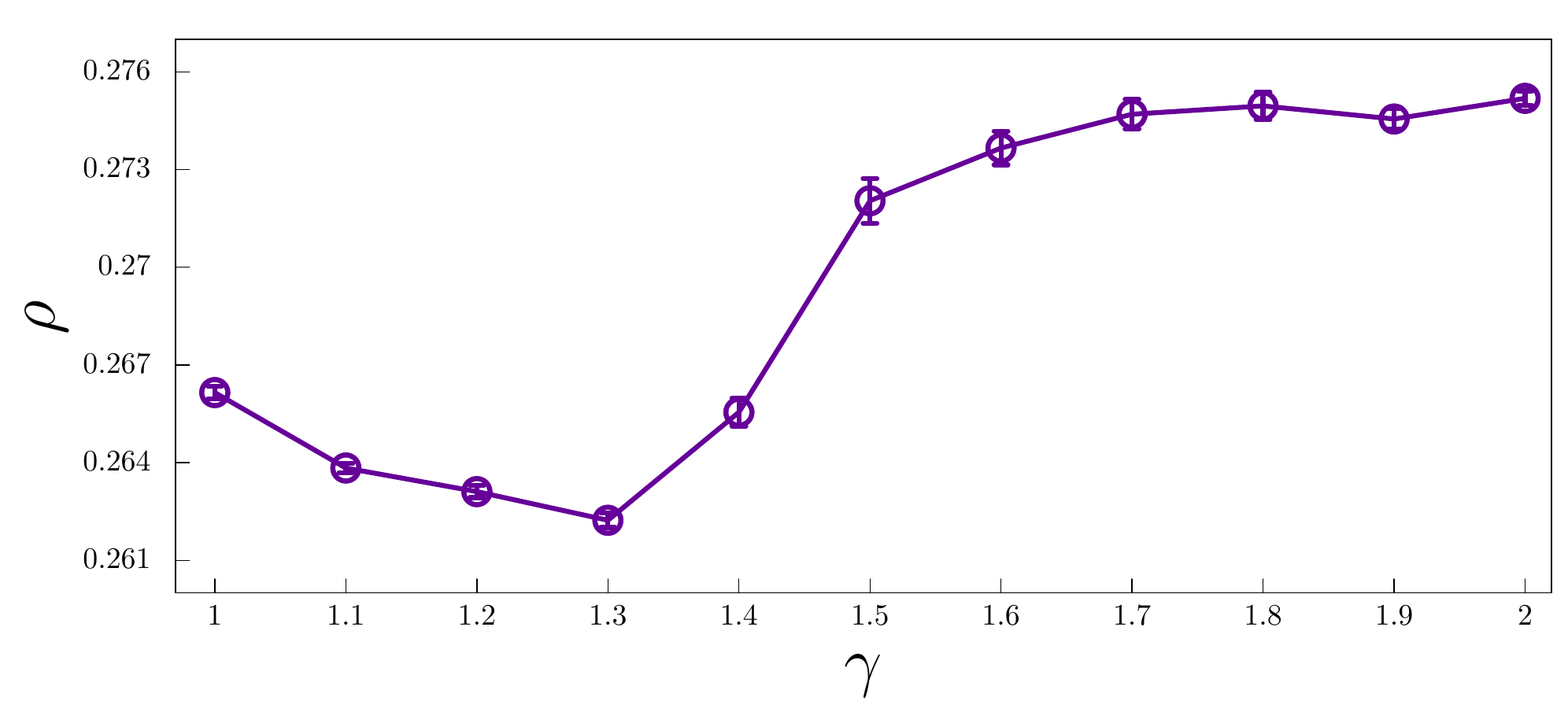}
\caption{Density of individuals of a single species in terms of the coinfection mortality catalysed factor. 
The outcomes were obtained by averaging the data from sets of $100$ simulations running in lattices with $500^2$ grid sites, running until $t=5000$ generations; the error bars show the standard deviation.
}
	\label{fig6}
\end{figure}

Figures \ref{fig5} and \ref{fig6} show the organisms' infection risk and species density dependence in the synergistic mortality factor $\gamma$.
The outcomes reveal that 
the organisms' infection risk
reaches its maximum if coinfection mortality increases in $30\%$.
Thus, the proportion of individuals dying due to complications of species $1$ or $2$ is the highest for $\gamma=1.3$, resulting in the minimum species density, as shown in Fig.~\ref{fig6}.
For $1.3 < \gamma <1.6$, $\chi$ is strongly reduced because of the reduction of the interface fluctuations, as shown in Fig.~\ref{fig3}. The variation of $\chi$ becomes then smooth for 
$\gamma >1.6$. Therefore the growth in the species densities is more significant for $1.3 < \gamma <1.6$.

\section{Discussion and Conclusions}
\label{sec7}
Our research addresses the spatial dynamics of two contagious disease epidemics spreading person-to-person in a cyclic model with three species.
As the pathogen transmission happens independently, they may infect the same organism at any time. 
When coinfection happens, there is a synergistic rise in the probability of death because of complications of one of the diseases.

Our stochastic simulations revealed the role played by the synergistic mortality factor in population dynamics.
If the diseases' severity is not significantly altered ($\gamma \approx 1$), both disease simultaneously affects almost all individuals in the grid, meaning that pathogens coexist, with the average number of coinfected individuals constant during the simulation.
As the diseases become more severe, individuals infected with only one of the pathogens live longer, creating departed single-disease spatial domains bordered by interfaces mainly formed by coinfected individuals. In this limit, the dynamics of the interface network are curvature-driven, with the total number of coinfected individuals approaching the following scaling law $I^{\star} \propto t^{-1/2}$. 

The dynamics of pathogens' spatial distribution 
impact organisms' safety, provoking a rise in infection
risk, reaching the maximum for an increase in disease severity of $30\%$. Because of this, the average densities of individuals of each species reaches the lowest for $\gamma=1.3$.
In contrast, for $\gamma>1.3$, coinfected individuals' synergistic death rate accelerates, weakening the viruses' transmission chain, thus, significantly reducing the organisms' infection risk. 
Therefore, we show evidence that, in a two-disease model, the extinction of one of the pathogens results from the topological aspects of two-dimensional space. The spontaneous symmetry breaking caused by the interactions' stochasticity defines how the synergistic mortality increase leads to eradicating one of the diseases. 

Our model can be generalised for more complex models of more than two diseases caused by distinct pathogens 
dominating departed spatial domains. We may assume different symmetries for 
the diseases' transmissibility and mortality parameters which result in diverse topological patterns in two-dimensional space simulations. As shown in other direct and apparent competition models, the pathogens may occupy patches forming separated by interfaces with $Y$-type junctions or string networks \cite{Avelino-PRE-86-031119,Pereira,apparent,Avelino-PRE-86-036112,PhysRevE.89.042710,PhysRevE.99.052310,strings1,strings2}. 
The outcomes may help ecologists to understand the dynamics of epidemics and the role of space topology in organisms' infection risk and disease eradication.
\section*{Acknowledgments}
We thank CNPq, ECT, Fapern, and IBED for financial and technical support.
\bibliographystyle{elsarticle-num}
\bibliography{ref}

\begin{thebibliography}{10}
\expandafter\ifx\csname url\endcsname\relax
  \def\url#1{\texttt{#1}}\fi
\expandafter\ifx\csname urlprefix\endcsname\relax\def\urlprefix{URL }\fi
\expandafter\ifx\csname href\endcsname\relax
  \def\href#1#2{#2} \def\path#1{#1}\fi

\bibitem{ecology}
M.~Begon, C.~R. Townsend, J.~L. Harper, Ecology: from individuals to
  ecosystems, Blackwell Publishing, Oxford, 2006.

\bibitem{Coli}
B.~Kerr, M.~A. Riley, M.~W. Feldman, B.~J.~M. Bohannan, Local dispersal
  promotes biodiversity in a real-life game of rock–paper–scissors, Nature
  418 (2002) 171.

\bibitem{Allelopathy}
R.~Durret, S.~Levin, Allelopathy in spatially distributed populations, J.
  Theor. Biol. 185 (1997) 165--171.

\bibitem{bacteria}
B.~C. Kirkup, M.~A. Riley, Antibiotic-mediated antagonism leads to a bacterial
  game of rock-paper-scissors in vivo, Nature 428 (2004) 412--414.

\bibitem{lizards}
B.~Sinervo, C.~M. Lively, The rock-scissors-paper game and the evolution of
  alternative male strategies, Nature 380 (1996) 240--243.

\bibitem{coral}
J.~B.~C. Jackson, L.~Buss, The rock-scissors-paper game and the evolution of
  alternative male strategies, Proc. Natl Acad. Sci. USA 72 (1975) 5160--5163.

\bibitem{mobiliahigh}
S.~Islam, A.~Mondal, M.~Mobilia, S.~Bhattacharyya, C.~Hens, Effect of mobility
  in the rock-paper-scissor dynamics with high mortality, Phys. Rev. E 105
  (2022) 014215.

\bibitem{social1}
E.~Du, E.~Chen, J.~Liu, C.~Zheng, How do social media and individual behaviors
  affect epidemic transmission and control?, Science of The Total Environment
  761 (2021) 144114.

\bibitem{disease4}
A.~M. Dunn, M.~E. Torchin, M.~J. Hatcher, P.~M. Kotanen, D.~M. Blumenthal,
  J.~E. Byers, C.~A. Coon, V.~M. Frankel, R.~D. Holt, R.~A. Hufbauer, A.~R.
  Kanarek, K.~A. Schierenbeck, L.~M. Wolfe, S.~E. Perkins, Indirect effects of
  parasites in invasions, Functional Ecology 26~(6) (2012) 1262--1274.

\bibitem{disease3}
M.~J. Young, N.~H. Fefferman, The dynamics of disease mediated invasions by
  hosts with immune reproductive tradeoff, Scientific Reports 12 (2022) 4108.

\bibitem{disease2}
T.~Nagatani, G.~Ichinose, K.~ichi Tainaka, Epidemics of random walkers in
  metapopulation model for complete, cycle, and star graphs, Journal of
  Theoretical Biology 450 (2018) 66--75.

\bibitem{socialdist}
T.~C. Reluga, Game theory of social distancing in response to an epidemic, PLoS
  Comput. Biol. 6~(5) (2010) e1000793.

\bibitem{soc}
S.~Stockmaier, N.~Stroeymeyt, S.~E. C., H.~D. M., L.~A. Meyers, D.~I. Bolnick,
  Infectious diseases and social distancing in nature, Science 371~(6533)
  (2021) eabc8881.

\bibitem{doi:10.1126/science.abc8881}
S.~Stockmaier, N.~Stroeymeyt, E.~C. Shattuck, D.~M. Hawley, L.~A. Meyers, D.~I.
  Bolnick, Infectious diseases and social distancing in nature, Science
  371~(6533)  eabc8881.

\bibitem{mr0}
G.~Dimarco, G.~Toscani, M.~Zanella, Optimal control of epidemic spreading in
  the presence of social heterogeneity, Philosophical Transactions of the Royal
  Society A: Mathematical, Physical and Engineering Sciences 380~(2224) (2022)
  20210160.

\bibitem{mr1}
Q.~Shao, D.~Han, Epidemic spreading in metapopulation networks with
  heterogeneous mobility rates, Applied Mathematics and Computation 412 (2022)
  126559.

\bibitem{mr2}
P.~Edsberg~Møllgaard, S.~Lehmann, L.~Alessandretti, Understanding components
  of mobility during the covid-19 pandemic, Philosophical Transactions of the
  Royal Society A: Mathematical, Physical and Engineering Sciences 380~(2214)
  (2022) 20210118.

\bibitem{10.1371/journal.pone.0254403}
T.~Oka, W.~Wei, D.~Zhu, The effect of human mobility restrictions on the
  covid-19 transmission network in china, PLOS ONE 16~(7) (2021) 1--16.

\bibitem{CAPAROGLU2021111246}
To restrict or not to restrict? use of artificial neural network to evaluate
  the effectiveness of mitigation policies: A case study of turkey, Chaos,
  Solitons \& Fractals 151 (2021) 111246.

\bibitem{plasticity2}
N.~Stroeymeyt, A.~V. Grasse, A.~Crespi, D.~P. Mersch, S.~Cremer, L.~Keller,
  Social network plasticity decreases disease transmission in a eusocial
  insect, Science 362~(6417) (2018) 941--945.

\bibitem{Gene}
V.~Papanikolaou, A.~Chrysovergis, V.~Ragos, E.~Tsiambas, S.~Katsinis,
  A.~Manoli, S.~Papouliakos, D.~Roukas, S.~Mastronikolis, D.~Peschos,
  A.~Batistatou, E.~Kyrodimos, N.~Mastronikolis, From delta to omicron:
  S1-rbd/s2 mutation/deletion equilibrium in sars-cov-2 defined variants, Gene
  814 (2022) 146134.

\bibitem{mutating1}
M.~Becerra-Flores, T.~Cardozo, Sars-cov-2 viral spike g614 mutation exhibits
  higher case fatality rate, International Journal of Clinical Practice 74~(8)
  (2020) e13525.

\bibitem{mutate2}
H.~M. Zawbaa, H.~Osama, A.~El-Gendy, H.~Saeed, H.~S. Harb, Y.~M. Madney,
  M.~Abdelrahman, M.~Mohsen, A.~M.~A. Ali, M.~Nicola, M.~O. Elgendy, I.~A.
  Ibrahim, M.~E.~A. Abdelrahim, Effect of mutation and vaccination on spread,
  severity, and mortality of covid-19 disease, Journal of Medical Virology
  94~(1) (2022) 197--204.

\bibitem{plasticity1}
A.~Bridier, P.~C. Piard, J-C.~and, S.~Labarthe, F.~Dubois-Brissonnet,
  R.~Briandet, Spatial organization plasticity as an adaptive driver of surface
  microbial communities, Front. Microbiol 8 (2017) 1364.

\bibitem{epidemicbook}
F.~M. Snowden, Epidemics and society: from the black death to the present, Yale
  University Press, New Haven and London, 2019.

\bibitem{tanimoto}
J.~Tanimoto, Sociophysics Approach to Epidemics, Springer,, Singapore, 2021.

\bibitem{epidemicprocess}
R.~Pastor-Satorras, C.~Castellano, P.~Van~Mieghem, A.~Vespignani, Epidemic
  processes in complex networks, Rev. Mod. Phys. 87 (2015) 925--979.

\bibitem{doi:10.1073/pnas.2007658117}
G.~Bonaccorsi, F.~Pierri, M.~Cinelli, A.~Flori, A.~Galeazzi, F.~Porcelli, A.~L.
  Schmidt, C.~M. Valensise, A.~Scala, W.~Quattrociocchi, F.~Pammolli, Economic
  and social consequences of human mobility restrictions under covid-19,
  Proceedings of the National Academy of Sciences 117~(27) (2020) 15530--15535.

\bibitem{combination}
E.~Rangel, B.~Moura, J.~Menezes, Combination of survival movement strategies in
  cyclic game systems during an epidemic, Biosystems 217 (2022) 104689.

\bibitem{jcomp}
J.~Menezes, B.~Moura, E.~Rangel, Adaptive survival movement strategy to local
  epidemic outbreaks in cyclic models, Journal of Physics: Complexity 3~(4)
  (2022) 045008.

\bibitem{plasticity}
Spatial organisation plasticity reduces disease infection risk in
  rock–paper–scissors models, Biosystems 221 (2022) 104777.

\bibitem{eplsick}
J.~Menezes, B.~Ferreira, E.~Rangel, B.~Moura, Adaptive altruistic strategy in
  cyclic models during an epidemic, Europhysics Letters 140~(5) (2022) 57001.

\bibitem{synergestic}
P.~Nicholson, N.~Mon-on, P.~Jaemwimol, P.~Tattiyapong, W.~Surachetpong,
  Coinfection of tilapia lake virus and aeromonas hydrophila synergistically
  increased mortality and worsened the disease severity in tilapia (oreochromis
  spp.), Aquaculture 520 (2020) 734746.

\bibitem{synergestic2}
M.~Chapwanya, A.~Matusse, Y.~Dumont, On synergistic co-infection in crop
  diseases. the case of the maize lethal necrosis disease, Applied Mathematical
  Modelling 90 (2021) 912--942.

\bibitem{synergestic3}
T.~Yan, S.~Zhu, H.~Wang, C.~Li, Y.~Diao, Y.~Tang, Synergistic pathogenicity in
  sequential coinfection with fowl adenovirus type 4 and avian orthoreovirus,
  Veterinary Microbiology 251 (2020) 108880.

\bibitem{synergestic4}
S.~TELFER, R.~BIRTLES, M.~BENNETT, X.~LAMBIN, S.~PATERSON, M.~BEGON, Parasite
  interactions in natural populations: insights from longitudinal data,
  Parasitology 135~(7) (2008) 767–781.

\bibitem{synergestic5}
M.~Singer, N.~Bulled, B.~Ostrach, E.~Mendenhall, Syndemics and the biosocial
  conception of health, The Lancet 389~(10072) (2017) 941--950.

\bibitem{synergestic8}
J.~C. Kash, K.-A. Walters, A.~S. Davis, A.~Sandouk, L.~M. Schwartzman, B.~W.
  Jagger, D.~S. Chertow, L.~Qi, R.~E. Kuestner, A.~Ozinsky, J.~K. Taubenberger,
  Lethal synergism of 2009 pandemic h1n1 influenza virus and streptococcus
  pneumoniae coinfection is associated with loss of murine lung repair
  responses, mBio 2~(5) (2011) e00172--11.

\bibitem{synergestic10}
C.~Yin, W.~Yang, J.~Meng, Y.~Lv, J.~Wang, B.~Huang, Co-infection of pseudomonas
  aeruginosa and stenotrophomonas maltophilia in hospitalised pneumonia
  patients has a synergic and significant impact on clinical outcomes, European
  Journal of Clinical Microbiology \& Infectious Diseases 36~(11) (2017)
  2231--2235.

\bibitem{virulence}
V.~Andreasen, A.~Pugliese, Pathogen coexistence induced by density-dependent
  host mortality, Journal of Theoretical Biology 177~(2) (1995) 159--165.

\bibitem{leonard}
R.~M. May, W.~J. Leonard, Nonlinear aspects of competition between three
  species, SIAM J. Appl. Math. 29 (1975) 243--253.

\bibitem{Avelino-PRE-86-031119}
P.~P. Avelino, D.~Bazeia, L.~Losano, J.~Menezes, von neummann's and related
  scaling laws in rock-paper-scissors-type games, Phys. Rev. E 86 (2012)
  031119.

\bibitem{Pereira}
T.~A. Pereira, J.~Menezes, L.~Losano, Interface networks in models of competing
  species, Intern. J. of Mod., Sim. and Sci. Comp. 9 (2018) 1850046.

\bibitem{apparent}
J.~Menezes, B.~Moura, Pattern formation and coarsening dynamics in apparent
  competition models, Chaos, Solitons \& Fractals 157 (2022) 111903.

\bibitem{Joana1}
P.~Avelino, J.~Oliveira, C.~Martins, Understanding domain wall network
  evolution, Physics Letters B 610~(1) (2005) 1--8.

\bibitem{Joana2}
P.~P. Avelino, R.~Menezes, J.~C. R.~E. Oliveira, Unified paradigm for interface
  dynamics, Phys. Rev. E 83 (2011) 011602.

\bibitem{Roman}
A.~Roman, D.~Konrad, M.~Pleimling, Cyclic competition of four species: domains
  and interfaces, J. Stat. Mech. 7 (2012) P07014.

\bibitem{Topics}
L.~F. Cugliandolo, Topics in coarsening phenomenai, Physica A 389 (2010)
  4360–4373.

\bibitem{Arezon1}
A.~Sicilia, J.~J. Arenzon, A.~J. Bray, L.~F. Cugliandolo, Domain growth
  morphology in curvature-driven two-dimensional coarsening, Phys. Rev. E 76
  (2007) 061116.

\bibitem{Interplay}
P.~Roy, P.~Sen, Interplay of interfacial noise and curvature-driven dynamics in
  two dimensions, Phys. Rev. E 95 (2017) 020101.

\bibitem{Arezon2}
M.~P.~O. Loureiro, J.~J. Arenzon, L.~F. Cugliandolo, A.~Sicilia,
  Curvature-driven coarsening in the two-dimensional potts model, Phys. Rev. E
  81 (2010) 021129.

\bibitem{Oliveira1}
M.~J. Oliveira, J.~F.~F. Mendes, M.~A. Santos, Nonequilibrium spin models with
  ising universal behaviour, J. Phys. A: Math. Gen. 26 (1993) 2317.

\bibitem{Tartaglia_2018}
A.~Tartaglia, L.~F. Cugliandolo, M.~Picco, Coarsening and percolation in the
  kinetic 2d ising model with spin exchange updates and the voter model
  2018~(8) (2018) 083202.

\bibitem{Avelino-PRE-86-036112}
P.~P. Avelino, D.~Bazeia, L.~Losano, J.~Menezes, B.~F. Oliveira, Junctions and
  spiral patterns in generalized rock-paper-scissors models, Phys. Rev. E 86
  (2012) 036112.

\bibitem{PhysRevE.89.042710}
P.~P. Avelino, D.~Bazeia, L.~Losano, J.~Menezes, B.~F. de~Oliveira, Interfaces
  with internal structures in generalized rock-paper-scissors models, Phys.
  Rev. E 89 (2014) 042710.

\bibitem{PhysRevE.99.052310}
P.~P. Avelino, J.~Menezes, B.~F. de~Oliveira, T.~A. Pereira, Expanding spatial
  domains and transient scaling regimes in populations with local cyclic
  competition, Phys. Rev. E 99 (2019) 052310.

\bibitem{strings1}
P.~Avelino, D.~Bazeia, J.~Menezes, B.~de~Oliveira, String networks in
  lotka–volterra competition models, Physics Letters A 378~(4) (2014) 393 --
  397.

\bibitem{strings2}
P.~Avelino, D.~Bazeia, L.~Losano, J.~Menezes, B.~de~Oliveira, String networks
  with junctions in competition models, Physics Letters A 381~(11) (2017)
  1014--1020.

\end{thebibliography}

\end{document}